\documentclass[11pt]{article}

\usepackage[preprint]{acl}
\usepackage{times}
\usepackage{latexsym}
\usepackage[T1]{fontenc}
\usepackage[utf8]{inputenc}
\usepackage{microtype}
\usepackage{inconsolata}
\usepackage{graphicx}
\usepackage{orcidlink}
\usepackage{amsmath}
\usepackage{amssymb}
\usepackage{booktabs}
\newcommand{\dataset}{\textsc{Real-2mix}}
\newcommand{\trainsetname}{\textsc{VoxBlink2-AVSE}}
\newcommand{\datarepositorylinks}{Repository links: \href{https://huggingface.co/datasets/phot0nqqq/Real2mix}{Real2mix} and \href{https://huggingface.co/datasets/phot0nqqq/Voxblink2_AVSE}{VoxBlink2-AVSE} on Hugging Face.}
\title{\dataset{}: Evaluating Identity Faithfulness in Audio-Visual Target Speaker Extraction}

\author{
  \textbf{Peijun Yang}\,\orcidlink{0009-0004-4948-2028}\textsuperscript{1,3},
  \textbf{Zhan Jin}\,\orcidlink{0009-0002-8157-7970}\textsuperscript{4},
  \textbf{Xiaoyi Qin}\,\textsuperscript{5},
  \textbf{Ruyi Gan}\,\textsuperscript{5},
  \textbf{Hao Wang}\,\textsuperscript{5},\\
  \textbf{Juan Liu}\,\orcidlink{0000-0001-9344-7415}\textsuperscript{3,4,*},
  \textbf{Ming Li}\,\orcidlink{0000-0002-6406-1983}\textsuperscript{2,3,*},
    \\
    \small{\textsuperscript{1}School of Cyber Science and Engineering, Wuhan University, Wuhan, China}
  \\
  \small{\textsuperscript{2}School of Artificial Intelligence, The Chinese University of Hong Kong, Shenzhen, China}
  \\
  \small{\textsuperscript{3}School of Artificial Intelligence, Wuhan University, Wuhan, China}
  \\
  \small{\textsuperscript{4}School of Computer Science, Wuhan University, Wuhan, China}
  \\
  \small{\textsuperscript{5}X Square Robot, Shenzhen, China}
  \\
  \small{\texttt{liujuan@whu.edu.cn, mingli369@cuhk.edu.cn}}
}

\hypersetup{pdftitle={Real-2mix: Evaluating Identity Faithfulness in Audio-Visual Target Speaker Extraction},pdfauthor={Peijun Yang, Zhan Jin, Xiaoyi Qin, Ruyi Gan, Hao Wang, Juan Liu, Ming Li},pdfsubject={Audio-visual target speaker extraction benchmark}}
\begin{document}
\maketitle
\begingroup
\renewcommand{\thefootnote}{*}
\footnotetext[1]{indicates the corresponding author.}
\endgroup

\begin{abstract}
Audio-visual target speaker extraction should recover the person indicated by the video, rather than repeatedly return the acoustically dominant voice. We introduce \dataset{}, a Mandarin benchmark of jointly recorded two-speaker mixtures with synchronized multi-view video and scene-matched solo references. Its 77 scenes contain 7,598 clips; filtering yields 6,038 evaluable mixtures, each queried for both speakers, giving 12,076 paired target queries. We also curate \trainsetname{}, comprising 250,828 synchronized audio--lip pairs. A seven-system TF-GridNet study examines visual representations, training corpora, and conditioning strategies. We evaluate transcription accuracy, speaker identity, residual overlap, and perceptual quality. The AVSR early-fusion system achieves 22.99\% CER, and 93.38\% of its outputs are more similar to the target speaker than to the interferer. Yet only 42.49\% satisfy both this identity criterion and the overlap detector's single-speaker criterion; both target queries satisfy these criteria for just 20.77\% of mixtures. A separate Real-AVSE challenge study examines development-remix adaptation. The benchmark reveals that favorable content and identity scores need not coincide with low detected overlap, motivating paired-target evaluation.
\end{abstract}

\section{Introduction}

Target speaker extraction (TSE) estimates one desired speaker from a mixture by conditioning a separator on a target cue. Audio-only systems commonly use a pre-recorded enrollment utterance \cite{zmolikova2019speakerbeam,zeng2025usef}.
Video provides a naturally synchronized alternative: visible lip motion identifies when a person speaks and supplies phonetic evidence that remains useful during voice overlap \cite{ochiai2019multimodal,pan2022usev}.
Audio-visual TSE is therefore attractive for meetings, interviews, and human--machine interaction. In these settings, speech recognition must preserve both what was said and who was selected: a fluent transcript of the interfering speaker answers the wrong target query.

Progress is nevertheless difficult to measure with mixtures created from independently recorded utterances. 
Such mixtures provide isolated sources for signal-reconstruction metrics, but they do not reproduce a shared physical scene in which target speech, interference, reverberation, and background noise are captured together. 
Real mixtures expose this synthetic-to-real gap, but usually lack isolated clean overlap sources for intrusive evaluation~\cite{li2025realt}. 
Moreover, a separator may favor one voice and weakly use its visual input. 
Evaluating each visual cue independently can miss this failure: two plausible-sounding outputs may both belong to the same person.

We introduce \dataset{} to benchmark identity-faithful AV-TSE on purpose-recorded real mixtures with synchronized audio and multi-view video.
Every A+B mixture is queried with both visual targets, 
while preceding A-only and B-only stages provide scene-matched voice references. 
Speaker similarity and CER then evaluate requested identity and linguistic content without requiring isolated clean waveforms for the overlap.
We complement this benchmark with \trainsetname, a large identity-disjoint audio-lip corpus curated for AVSE training,
and with a seven-system study of frozen visual representations and fusion strategies in TF-GridNet. We also examine real-remix adaptation using AV-SpatialNet under the Real-AVSE challenge protocol.

Our contributions are threefold:
\begin{itemize}
    \item We introduce \dataset{}, a real-scene benchmark with jointly recorded overlap, 28 speakers in 14 fixed dyads, scene-matched solo references, and synchronized multi-view video.
    \item We evaluate both visual targets of each mixture, combining content recovery, relative speaker identity, and overlap diagnostics to reveal limitations of single-output assessment.
    \item We curate \trainsetname{}, with 250,828 audio--lip pairs and speaker-disjoint train/dev partitions, and study seven TF-GridNet configurations. Fusion gains depend on the visual representation; a separate Real-AVSE adaptation study provides complementary evidence.
\end{itemize}

\section{Related Work}

\subsection{Speech Separation and Target Extraction}

Conv-TasNet established a strong time-domain speech-separation baseline \cite{luo2019convtasnet}, while TF-GridNet combines frequency and temporal modeling in the time-frequency domain \cite{wang2023tfgridnet}.
Unlike permutation-invariant separation, TSE uses an auxiliary cue to specify which source to extract.
SpeakerBeam uses an enrollment utterance \cite{zmolikova2019speakerbeam}, whereas Multimodal SpeakerBeam incorporates audio and visual clues \cite{ochiai2019multimodal}.
VisualVoice imposes cross-modal consistency \cite{gao2021visualvoice}, and USEV addresses mixtures ranging from no overlap to complete overlap \cite{pan2022usev}. AV-GridNet further grounds TF-GridNet with a target face recording \cite{pan2023avgridnet}.
Our focus is complementary: we benchmark AV-TSE on jointly recorded real mixtures and test whether changing the visual target changes the extracted identity, rather than only whether one prompted output sounds clean.

\subsection{Visual Speech Representations and Corpora}

AV-HuBERT learns audio-visual speech representations through masked multimodal cluster prediction \cite{shi2022avhubert}.
M2S-AVSR extends this line toward multi-view and real-scene robustness \cite{su2026m2savsr}.
We use the visual-only AV-HuBERT branch and the learned linear projection from the latter implementation, rather than its ASR decoder.

VoxCeleb and VoxBlink2 collect large-scale in-the-wild face--voice material for speaker recognition \cite{nagrani2017voxceleb,lin2024voxblink2}, LRS2 and LRS3 supply sentence-level material for visual speech recognition \cite{chung2017lrs2,afouras2018lrs3}; and AVSpeech enables audio-visual mixtures at scale \cite{ephrat2018looking}.
These corpora provide valuable pretraining material, but they do not directly benchmark AV-TSE on jointly captured overlap, where target speech, interference, reverberation, and background noise are recorded through the same physical scene.
REAL-T exposes the corresponding synthetic-to-real gap for audio-only TSE by constructing test mixtures from conversational recordings and evaluating ASR error because isolated clean sources are unavailable~\cite{li2025realt}.
\dataset{} targets this gap for AV-TSE with synchronized audio, multi-view video, and real two-speaker mixtures.
Its A-only and B-only stages provide in-scene enrollment references; the benchmark centers on extracting either speaker from the same recorded A+B mixture and evaluates the outputs with speaker similarity and CER, without requiring isolated clean waveforms for the overlap.
\trainsetname{} complements this real-mixture benchmark with quality-controlled, aligned audio-lip pairs for AVTSE training.
\begin{figure*}[t]
    \centering
    \includegraphics[width=\textwidth]{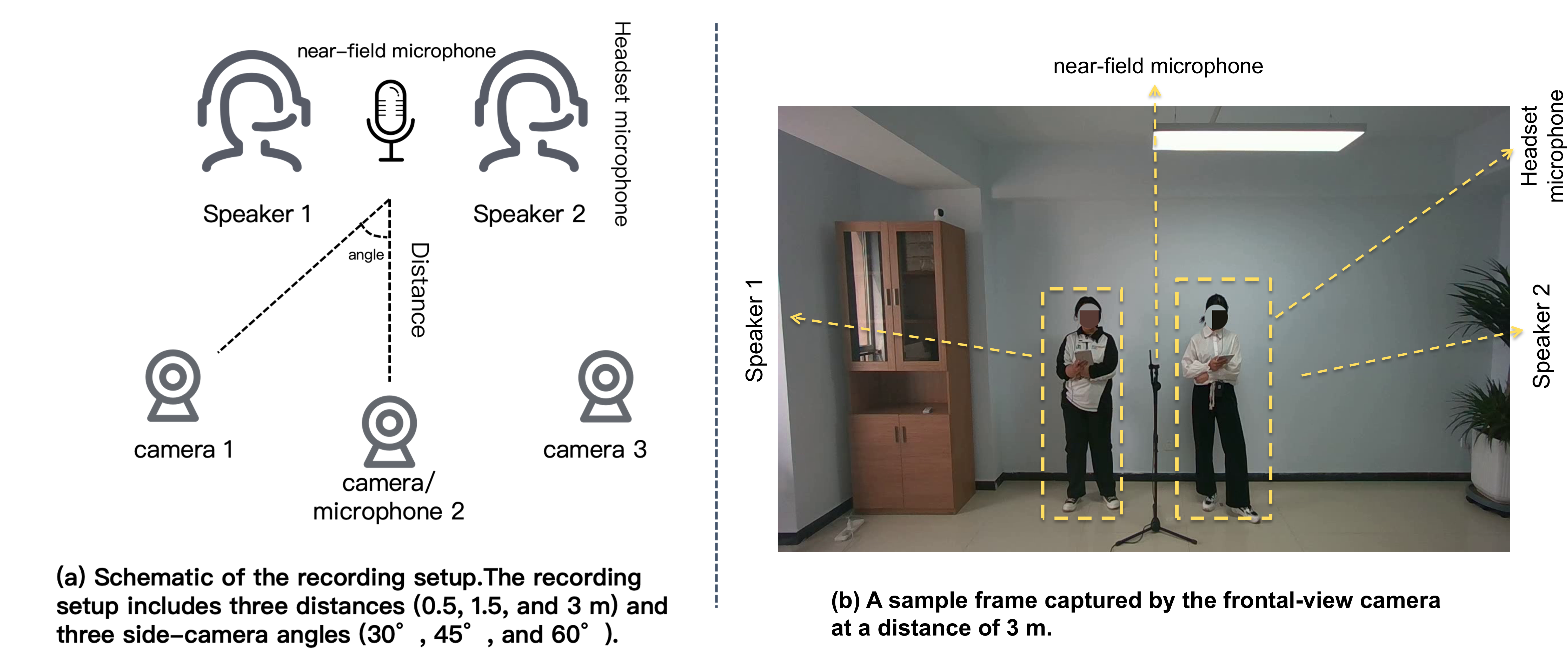}
    \caption{\dataset{} recording setup and a representative frontal-view frame. Panel (a) shows the camera geometry and microphone arrangement; panel (b) shows a 3-m far-field example with the two speaker regions marked by dashed boxes.}
    \label{fig:Real-2mix-recording}
\end{figure*}

\section{The \dataset{} Benchmark}

\subsection{Task Definition and Recording Protocol}

Let an overlap recording be $x(t)=(h_A*u_A)(t)+(h_B*u_B)(t)+n(t)$, where $u_A,u_B$ are source speech signals, $h_A,h_B$ are full room impulse responses (RIRs), $*$ denotes convolution, and $n$ is background noise. Given the visual stream $v_q$ for target $q\in\{A,B\}$, the extractor predicts
\begin{equation}
    \hat{s}_q=f_\theta(x,v_q).
\end{equation}
The desired target contains direct speech and early reflections: training uses the first 50 ms of the target RIR (Section~\ref{sec:method}), rather than the full reverberant contribution in $x$. Paired queries should recover A's speech with $v_A$ and B's with $v_B$, instead of returning the same dominant voice for both cues.

Two stationary participants record A-only, B-only, and overlapping reading stages under fixed conditions. Solo stages supply scene-matched references; one reader starts during the other's speech to create natural overlap. Speech is read Mandarin, excluding non-verbal vocalizations. Full-length recordings are manually segmented, aiming to retain a complete utterance from one speaker at each boundary; the other speaker need not finish simultaneously.

\subsection{Geometry and Acquisition}

Figure~\ref{fig:Real-2mix-recording} shows a central camera and two symmetric side cameras at $30^{\circ}$, $45^{\circ}$, or $60^{\circ}$ relative to the participant midpoint, with nominal distances 0.5, 1.5, or 3 m. Eye-level cameras retain both mouths. Camera-synchronized audio, used for the benchmark, is 16 kHz; headset and near-field recordings are 48 kHz. All three cameras record the same two-person speaking event, so viewpoints are not independent events. Evaluation uses predominantly frontal views, with some non-frontal views retained. Appendix~\ref{sec:corpus-details} gives acquisition details.

\subsection{Corpus Composition and Availability}

The corpus comprises 77 independently recorded scenes with 28 distinct speakers in 14 fixed dyads; no speaker belongs to more than one dyad.
Its 7,598 mono 16-kHz clips total 11.84 hours.
Of these clips, 6,042 are dual-speaker mixtures from Stage III, while 1,556 contain solo speech from Stage I or II.
The $30^{\circ}$, $45^{\circ}$, and $60^{\circ}$ settings contain 30, 23, and 24 recording scenes, respectively.

Of the 6,042 mixtures with audio and both speaker transcripts, 6,038 have usable aligned visual inputs across all 77 scenes.
Each usable mixture appears twice in the manifest, once per visual target, yielding 12,076 target-speaker queries rather than 12,076 independent recordings.

A subset of the recordings is adopted by the Real-World AVSE Challenge~\cite{li2026realavse}; Section~\ref{sec:real-avse} distinguishes its test protocol from the full benchmark. A professional data-collection company recorded \dataset{}.\footnote{\datarepositorylinks} All recruited participants signed consent forms covering data collection and public release. \dataset{} is released under CC BY-NC 4.0.

\begin{table}[t]
\centering
\caption{Complete \dataset{} statistics by nominal distance. Simultaneous views share an event. Hours are rounded independently; the total is computed before rounding.}
\label{tab:Real-2mix-stats}
\begin{tabular}{lrrr}
\toprule
Distance & Scenes & Clips & Hours \\
\midrule
0.5 m & 22 & 1,987 & 3.22 \\
1.5 m & 40 & 4,225 & 6.36 \\
3 m & 15 & 1,386 & 2.25 \\
\midrule
Total & 77 & 7,598 & 11.84 \\
\bottomrule
\end{tabular}
\end{table}

\subsection{Cross-Video Identity and Speaker Alignment}

RetinaFace \cite{deng2020retinaface} and a Face Alignment Network \cite{bulat2017facealignment} detect and stabilize two faces, after which landmarks 48--67 define grayscale $96\times96$ mouth ROIs.
Frame-local tracks can exchange identities, so we cluster FaceNet-style embeddings \cite{schroff2015facenet,cao2018vggface2} into two scene-level prototypes and use Viterbi decoding~\cite{viterbi1967error} to obtain a temporally consistent keep-or-swap sequence.
TalkNet \cite{tao2021talknet} activity scores on the 1,556 solo clips then align the two stable visual identities with transcript labels s1 and s2; uncertain scenes are manually reviewed.
Appendix \ref{sec:identity-alignment-details} gives the objective and decision details.

Of 7,598 recorded clips, 7,594 videos enter the decoded-video processing set, totaling 1,064,818 frames. The four remaining clips lack usable ROI pairs and are excluded from paired-query evaluation.
It corrects identity exchanges in 52 videos (15,594 reassigned frames, 1.46\%) and repairs approximately 32,831 duplicate-detection or collision frames.
Evaluation requires audio, both transcripts, and both aligned mouth ROIs. Each mixture is queried twice using the fixed person-to-speaker mapping, without test-time permutation.

\section{The \trainsetname{} Training Corpus}

We curate synchronized audio--lip pairs from VoxBlink2~\cite{lin2024voxblink2}, whose source release contains 9.90 million utterances from 111,284 identities. Each retained item has decodable 16-kHz mono speech, a synchronized grayscale $T\times96\times96$ mouth track, and a stable identity--video--segment key. Audio-visual validation removes unusable or unaligned pairs; subsequent DNSMOS~\cite{reddy2021dnsmos} filtering retains audio with predicted overall quality $\mathrm{OVRL}\geq3.3$. We discard rejected pairs rather than enhance them, preserving synchronization without introducing denoising artifacts. This threshold favors higher predicted quality at the cost of corpus size, while retained recordings can still contain noise and reverberation.

The resulting corpus contains 250,828 pairs from 28,421 identities, totaling 766.17 hours, with mean segment duration 10.99 seconds and mean OVRL 3.38. The speaker-disjoint training and development partitions contain 239,680/11,148 pairs from 26,927/1,494 identities, respectively. Curation retains 25.54\% of source identities but only 2.53\% of utterances, emphasizing broad identity coverage over many clips per speaker. We use \dataset{} only for real-scene testing.

\trainsetname{} releases file lists, speaker-disjoint manifests, and curation instructions under CC BY-NC 4.0. Users obtain VoxBlink2 under its upstream terms and extract the listed audio/ROI pairs; the release does not redistribute or relicense VoxBlink2 media. Appendix~\ref{sec:corpus-details} details curation and compares other audio-visual corpora.

\section{Visual Conditioning in TF-GridNet}
\label{sec:method}

We study seven baseline configurations under a common six-block, single-output TF-GridNet backbone~\cite{wang2023tfgridnet}. The study varies the frozen visual representation, separation-training corpus, visual-fusion strategy, and an auxiliary speaker loss. All seven systems are noncausal. Fusion strategies provide two contrasting ways to condition the benchmark systems on visual input. Early fusion (F1 and X1) adds a visual frequency map to the mixture input, whereas the adapter systems (B0, L1, B1, D1, and D2) inject visual-conditioned residuals into each intra-frequency branch without an input visual map.

\begin{figure*}[t]
\centering
\includegraphics[width=\textwidth]{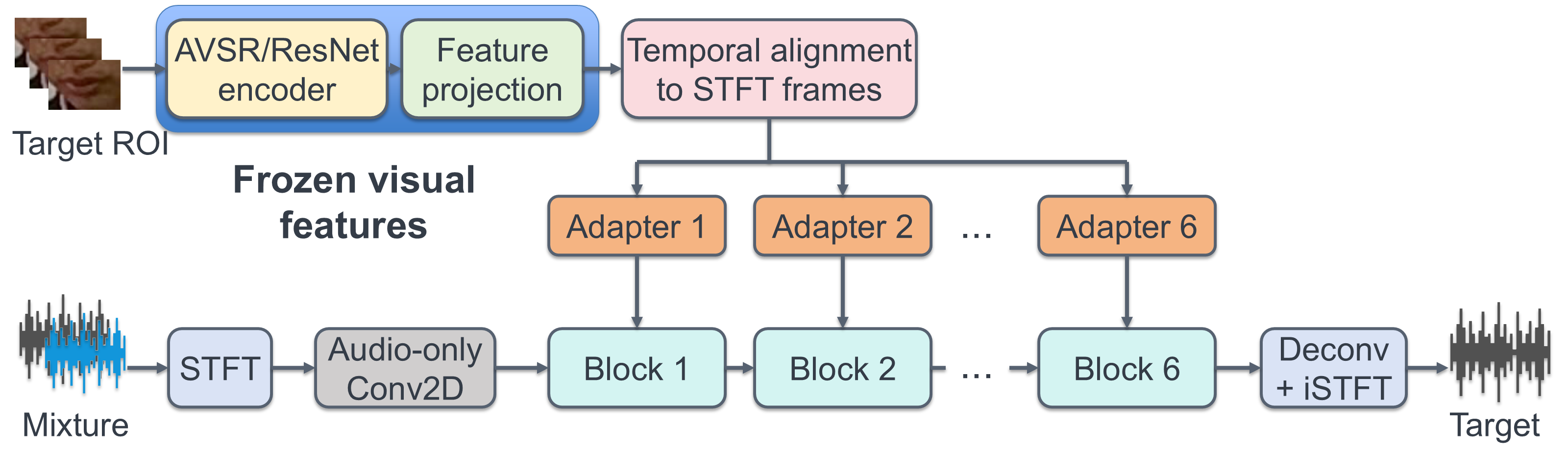}
\caption{Adapter conditioning, one of the two fusion strategies studied. Only mixture STFT components enter the input convolution. Aligned visual features and normalized audio activations jointly form residuals inside each intra-frequency branch of six GridNet blocks. Blocks 3--5 and internal audio-to-adapter connections are omitted for clarity. Early fusion is a separate configuration (Section~\ref{sec:early-fusion}).}
\label{fig:model-overview}
\end{figure*}

\subsection{Frozen Visual Representations}

\textbf{AVSR1280.} We use the video branch and learned projection of the AV-HuBERT-based encoder used by M2S-AVSR~\cite{su2026m2savsr}. The mixture audio is not provided to this feature extractor. The offline pipeline repairs invalid frames, center-crops mouth ROIs to $88\times88$, and applies the encoder's input normalization. A 1,024-dimensional backbone output is projected to 1,280 dimensions. The backbone and projection are frozen during separator training, and features are precomputed at 25 frames/s. The visual frontend is noncausal.

\textbf{ResNet512.} We also use 512-dimensional features from a frozen VisualFeatNet ResNet backbone. B1, X1, and D2 share the same precomputed test-feature manifest. The comparison evaluates each representation as a complete frontend, including its backbone, pretraining, and dimensionality. Both representations are linearly interpolated to the audio STFT frame rate with \texttt{align\_corners=False}.

\subsection{Common Separation Backbone}

Audio is mono at 16 kHz, standardized before separation and rescaled afterward. A centered 128-sample Hann STFT with hop 64 produces 65 bins. Six 128-channel GridNet blocks combine bidirectional intra-frequency/inter-time LSTMs and four-head unmasked temporal attention. A convolutional decoder predicts the complex target spectrum for iSTFT. The system is noncausal despite its 8-ms STFT window. Appendix~\ref{sec:adapter-details} specifies backbone and adapter operations.

\subsection{Input-Only Early Fusion}
\label{sec:early-fusion}

For F1 and X1, a visual-to-frequency network maps the aligned $V$-dimensional visual sequence through a $1\times1$ projection to 512 channels, GELU, a temporal convolutional subnetwork, and a final projection to 65 frequency bins. The resulting map $g(v)\in\mathbb{R}^{B\times1\times T\times65}$ is concatenated with the real and imaginary mixture components:
\begin{equation}
 h_0=\mathrm{GN}\bigl(\mathrm{Conv}_{3\times3}[\Re X,\Im X,g(v)]\bigr).
\end{equation}
No additional visual residual is injected into the GridNet blocks in this path.

\subsection{Layer-Wise Low-Rank AV Adapter}

The adapter path instead uses only $[\Re X,\Im X]$ at the input convolution. In block $\ell$, an independent adapter augments the intra-frequency residual:
\begin{align}
 z_\ell &= \mathrm{intra\_norm}(x_\ell),\\
 x'_\ell &= x_\ell+F_{\mathrm{intra},\ell}(z_\ell)+U_\ell(z_\ell,v).
\end{align}
The inter-time branch and self-attention then process $x'_\ell$. Unlike early fusion, this residual interface leaves the audio input channel count unchanged and separates condition injection from input encoding. The interface supports modular conditioning within the separator, with the adapter learned jointly during training.

The rank-32 adapter combines projected audio and visual features through a multiplicative interaction, then adds a frame-wise RMS-bounded residual. Its learned scalar strength starts at zero, so conditioning is introduced gradually during optimization. Appendix~\ref{sec:adapter-details} gives the projection, normalization, and scaling equations. Together, these design choices define the internal-conditioning strategy compared with early fusion.

\subsection{Separation and Speaker Objectives}

Online mixtures combine independently sampled target and interferer utterances convolved with sampled RIRs. The mixture uses the full responses, while the supervised target $s_{\mathrm{early}}$ uses the first 50 ms of the target RIR. The supervised target therefore retains early reflections. Each example is supervised once with its designated target visual cue; there is no paired A/B consistency loss.

We write SI-SNR with the estimate first and the reference second. Six systems minimize $\mathcal{L}_{\mathrm{sep}}=-\mathrm{SI\mbox{-}SNR}(\hat{s},s_{\mathrm{early}})$. Only L1 adds a speaker-margin term:
\begin{align}
 \mathcal{L}_{\mathrm{spk}} &={\mathbb E}\,[0.2+\cos(\hat e,e_i)-\cos(\hat e,e_t)]_+,\\
 \mathcal{L}_{\mathrm{L1}} &= \mathcal{L}_{\mathrm{sep}}+0.1\mathcal{L}_{\mathrm{spk}}.
\end{align}
The training speaker encoder is a frozen SpeakerLab ECAPA-TDNN~\cite{desplanques2020ecapa} trained on CNCeleb, with 192-dimensional embeddings. Its output-audio path permits gradients to the separator, whereas reference embeddings are detached. Systems without this loss do not execute the auxiliary encoder.

\section{Experiments}
\label{sec:experiments}

\subsection{Numbered Configurations and Training Data}

Table~\ref{tab:numbered-configs} summarizes the seven configurations and their real-scene results at the selected checkpoints. All seven use the six-block backbone in Section~\ref{sec:method}. Real-AVSE is reported separately.

\begin{table*}[t]
\centering\small
\setlength{\tabcolsep}{5pt}
\caption{Configurations and core results for the seven numbered systems on 12,076 target outputs. VB2/VC2 denote VoxBlink2-AVSE/VoxCeleb2 training data; $\lambda_{\mathrm{spk}}$ is the auxiliary speaker-loss weight. CER and $P(m>0)$ are percentages; Margin is mean target-minus-interferer cosine similarity. Positive margin is an identity-preference proxy, not verified extraction success. Bold marks the best result among these systems.}
\label{tab:numbered-configs}
\begin{tabular}{@{}llllrrrr@{}}
\toprule
ID & Train & Visual & Fusion & $\lambda_{\mathrm{spk}}$ & CER $\downarrow$ & Margin $\uparrow$ & $P(m>0)$ $\uparrow$ \\
\midrule
B0 & VB2 & AVSR1280 & Adapter & 0 & 23.66 & 0.3152 & 90.68 \\
L1 & VB2 & AVSR1280 & Adapter & 0.1 & 24.99 & 0.3131 & 89.30 \\
F1 & VB2 & AVSR1280 & Early fusion & 0 & \textbf{22.99} & \textbf{0.3428} & \textbf{93.38} \\
B1 & VB2 & ResNet512 & Adapter & 0 & 36.57 & 0.2164 & 78.83 \\
D1 & VC2 & AVSR1280 & Adapter & 0 & 34.98 & 0.2551 & 81.07 \\
X1 & VB2 & ResNet512 & Early fusion & 0 & 38.10 & 0.2060 & 78.02 \\
D2 & VC2 & ResNet512 & Adapter & 0 & 39.00 & 0.2020 & 75.52 \\
\bottomrule
\end{tabular}
\end{table*}

The VoxBlink2 training partition contains 239,680 paired utterances from 26,927 identities. D1 and D2 instead use a VoxCeleb2~\cite{chung2018voxceleb2} subset containing 232,266 utterances from 1,256 identities. Thus the training item counts differ by only 3.19\%, while speaker coverage differs substantially. The subsets also differ in recording domain, acoustic quality, and utterance duration, alongside speaker diversity. All seven systems, including D1 and D2, select checkpoints using the same VoxBlink2 development partition of 11,148 utterances from 1,494 identities. This common validation domain is a potential confound in the training-corpus comparison.

\subsection{Optimization and Checkpoint Selection}

We train from scratch on 3-second, 16-kHz segments using online RIR mixtures and no separate background-noise source. The nominal SIR draw is $\max(\mathcal{N}(5,10^2),-5)$ dB; the gain uses early-target and full-interferer power, so realized mixture SIR also depends on reverberation. Development mixtures are also sampled online.

All runs use FP32 AdamW for 50 epochs, backbone/visual learning rates $10^{-3}/10^{-4}$, gradient clipping at 5, and normally 8 GPUs $\times$ 6 examples. Checkpoint selection uses development loss; \dataset{} is reserved for evaluation. Appendix~\ref{sec:training-provenance} provides optimizer settings and additional implementation details.

\begin{table*}[t]
\centering\small
\setlength{\tabcolsep}{5pt}
\caption{Overlap--identity proxies and DNSMOS P.835 scores. Overlap, JointPass, and Wrong are mutually exclusive percentages over 12,076 outputs; BothID and BothPass are percentages over 6,038 mixtures. JointPass and BothPass are automatic proxy metrics, not ground-truth separation success. DNSMOS scores are not percentages.}
\label{tab:osd-strict-results}
\begin{tabular}{@{}lrrrrrrrr@{}}
\toprule
ID & Overlap $\downarrow$ & JointPass $\uparrow$ & Wrong $\downarrow$ & BothID $\uparrow$ & BothPass $\uparrow$ & SIG $\uparrow$ & BAK $\uparrow$ & OVRL $\uparrow$ \\
\midrule
B0 & 75.11 & 22.92 & \textbf{1.97} & 82.81 & 5.58 & \textbf{2.7993} & 2.8271 & 2.1194 \\
L1 & 65.61 & 31.53 & 2.87 & 79.94 & 10.20 & 2.7004 & 2.9305 & 2.0995 \\
F1 & 55.45 & 42.49 & 2.06 & \textbf{87.91} & 20.77 & 2.7314 & \textbf{2.9429} & \textbf{2.1270} \\
B1 & 72.52 & 23.49 & 3.98 & 59.19 & 5.68 & 2.6386 & 2.7817 & 1.9892 \\
D1 & \textbf{44.85} & \textbf{48.37} & 6.78 & 64.77 & \textbf{26.55} & 2.4365 & 2.1049 & 1.6602 \\
X1 & 80.52 & 16.09 & 3.39 & 58.46 & 3.23 & 2.6537 & 2.6529 & 1.9526 \\
D2 & 85.62 & 11.73 & 2.65 & 55.78 & 2.27 & 2.5732 & 2.4683 & 1.8352 \\
\bottomrule
\end{tabular}
\end{table*}

\subsection{Complete Real-Scene Evaluation}

Every one of the 6,038 retained mixtures is evaluated twice, using A's and B's visual cues, for 12,076 target-speaker outputs across 77 scenes. Each experiment covers the complete output set. Inference is full-utterance FP32 with batch size one, using the fixed physical person0/s1 and person1/s2 mapping, without test-time permutation, 3-second truncation, or cross-length padded batches. The visual input is the corresponding frozen offline representation, and no speaker enrollment is supplied to the separator. Solo-stage recordings are used only by the identity evaluator.

\paragraph{Content: CER.}
We transcribe each output with Qwen3-ASR-1.7B~\cite{shi2026qwen3asr}, using Chinese mode, FP16, batch size one, and a maximum of 256 new tokens. Reference and hypothesis undergo the same predefined Chinese/ASCII punctuation removal and whitespace normalization. The micro-averaged character error rate is
\begin{equation}
 \mathrm{CER}=\frac{\sum_{i,q}\mathrm{ED}(\mathcal{A}(\hat{s}_{i,q}),y_{i,q})}{\sum_{i,q}|y_{i,q}|},
\end{equation}
where $\mathrm{ED}$ is Levenshtein distance~\cite{levenshtein1966binary}. The denominator is 211,866 reference characters, and empty transcriptions still contribute errors. Tables report CER as a percentage, not the mean of utterance-level CERs.

\paragraph{Identity: speaker similarity.}
The evaluator is the official WeSpeaker ResNet34 ONNX model trained on VoxCeleb~\cite{wang2022wespeaker}. For each scene and speaker, we average up to five L2-normalized solo-stage embeddings and normalize the mean again. Table~\ref{tab:numbered-configs} reports the margin $m=S_t-S_i$ and $P(m>0)$; target/interferer cosine similarities $S_t$/$S_i$ appear in Table~\ref{tab:numbered-details}. A positive margin indicates relative preference for the target; it can also occur in outputs containing both speakers. The evaluator differs from L1's auxiliary encoder in architecture, implementation, and pretraining data; potential identity overlap with the VoxCeleb2 separation-training subset remains unaudited.

\paragraph{Joint assessment of speaker identity and residual overlap.}
We combine speaker margin with residual-overlap detection to assess whether an output favors the intended speaker while containing little detected overlap. The fixed WavLM classifier is distributed as \path{Den4ikAI/speech_overlap_detection}. Only the OSD input is normalized to RMS 0.05. Overlap posteriors are averaged over valid 2-second windows with a 0.5-second hop. With threshold $\tau=0.35$, we report
\begin{align}
 \mathrm{Overlap} &= P(\bar p\geq\tau),\\
 \mathrm{JointPass} &= P(\bar p<\tau\ \land\ m>0),\\
 \mathrm{Wrong} &= P(\bar p<\tau\ \land\ m\leq0).
\end{align}
These categories partition the outputs: those above the overlap threshold enter Overlap regardless of speaker margin, and the remaining outputs are divided by identity preference. JointPass combines both automatic criteria. BothPass requires JointPass for both targets of one mixture and uses 6,038 mixtures as its denominator. We also report both-target positive margin (BothID). The OSD threshold was chosen during prior exploratory listening; independent calibration and sensitivity analysis remain open. Appendix~\ref{sec:metric-details} documents edge cases.

\paragraph{Perceived quality: DNSMOS.}
We report the official non-personalized DNSMOS P.835 SIG, BAK, and OVRL scores~\cite{reddy2021dnsmosp835}. OSD-specific RMS normalization is not passed to this evaluator. These automatic perceptual-quality estimates complement the content and identity diagnostics.

Evaluation combines transcript recovery, relative speaker identity, detected overlap, and perceptual quality. The preceding solo stages provide scene-matched identity references recorded separately from the jointly spoken mixtures.

\subsection{Results and Analysis}
\label{sec:ablations}

\paragraph{Visual representation.}
The AVSR representation lowers CER in all three corresponding comparisons: B1$\rightarrow$B0 by 12.91 percentage points on VoxBlink2 with adapters; X1$\rightarrow$F1 by 15.11 points with early fusion; and D2$\rightarrow$D1 by 4.02 points on VoxCeleb2 with adapters. The consistent CER reductions favor the selected AVSR representation for content recovery. The comparison evaluates the representation as a whole, including its backbone, dimensionality, projection, and pretraining objective.

\paragraph{Separation-training corpus.}
With AVSR adapters, D1$\rightarrow$B0 reduces CER by 11.32 points; with ResNet adapters, D2$\rightarrow$B1 reduces it by 2.43 points. Speaker margin also improves in both comparisons. However, B0 has more OSD-detected residual overlap and lower JointPass than D1. The corpus comparison shows a trade-off between content recovery and overlap diagnostics, with speaker coverage, domain, acoustic quality, and the shared validation set varying together.

\begin{table*}[t]
\centering
\footnotesize
\setlength{\tabcolsep}{2.5pt}
\caption{Real-AVSE Track~1 real-mixture test comparison. Official Track~1 ranks cover both mix and remix; ``=1'' denotes joint first. CER is percent ($\downarrow$); other scores are $\uparrow$. Bold marks the best score in each column.}
\label{tab:real-avse-results}
\begin{tabular}{@{}lcrrrrrrr@{}}
\toprule
System & Rank & UTMOS & DNS-P808 & DNS-SIG & DNS-BAK & DNS-OVRL & CER (\%) & SPK-SIM \\
\midrule
audioman~\cite{li2026realavse} & =1 & 2.0243 & 2.9502 & 2.3016 & 2.5344 & 1.8022 & 14.72 & 0.7336 \\
twilight~\cite{li2026realavse} & 3 & \textbf{2.1814} & 2.9240 & 2.5005 & 2.6145 & 2.0980 & 15.79 & 0.7256 \\
AITD~\cite{li2026realavse} & =1 & 2.0716 & 2.7369 & 2.1091 & 2.3253 & 1.6641 & 18.32 & \textbf{0.7511} \\
\midrule
SpatialNet (ours) & -- & 1.4062 & 2.9354 & \textbf{2.6588} & 3.0567 & 2.1285 & 16.76 & 0.5148 \\
SpatialNet-FT (ours) & -- & 1.8616 & \textbf{3.3012} & 2.5452 & \textbf{3.2641} & \textbf{2.1370} & \textbf{12.80} & 0.6381 \\
\bottomrule
\end{tabular}
\par\smallskip
\begin{minipage}{\textwidth}
\footnotesize
The three listed competing systems also use development-remix fine-tuning; adaptation recipes vary by system. SpatialNet uses simulated pretraining only; SpatialNet-FT adds development-remix fine-tuning. ``--'' indicates no verified official rank.
\end{minipage}
\end{table*}

\paragraph{Fusion depends on the representation.}
With ResNet512, replacing early fusion X1 by adapter B1 lowers CER from 38.10\% to 36.57\%, reduces residual overlap from 80.52\% to 72.52\%, and raises JointPass from 16.09\% to 23.49\%. Target similarity nevertheless decreases from 0.5622 to 0.5486. With AVSR1280, early fusion F1 is better than adapter B0 in CER (22.99\% versus 23.66\%), residual overlap (55.45\% versus 75.11\%), and JointPass (42.49\% versus 22.92\%). Their OVRL scores are close (2.1270 versus 2.1194), whereas B0 has higher SIG. Thus adapters improve several metrics with ResNet features, but early fusion is stronger with AVSR features. Their residual interface provides modular conditioning without changing input channels.

Adapters have fewer separator parameters (Table~\ref{tab:numbered-details}): B0 uses 10,070,102 versus F1's 11,516,178, a 12.56\% reduction; B1 uses 9,922,646 versus X1's 11,122,962, a 10.79\% reduction. Counts cover the separator, excluding frozen visual frontends and external speaker models.

\paragraph{The auxiliary-loss configuration exhibits a trade-off.}
B0$\rightarrow$L1 increases CER from 23.66\% to 24.99\% and lowers the positive-margin fraction, but reduces residual overlap from 75.11\% to 65.61\% and improves JointPass from 22.92\% to 31.53\%. Thus L1 exhibits a trade-off rather than uniformly better extraction.

\paragraph{Content, identity, and residual overlap.}
F1 has the lowest CER, highest target similarity, largest speaker margin, and highest positive-margin fraction among these seven checkpoints. Its 93.38\% positive-margin rate contrasts with only 42.49\% JointPass and 20.77\% BothPass, showing that relative identity preference and the overlap diagnostic need not agree. D1 has the highest JointPass (48.37\%) and BothPass (26.55\%) but worse CER (34.98\%) and OVRL (1.6602). The ranking therefore depends on whether content, relative identity, detected overlap, or perceptual quality is emphasized.

\subsection{Real-AVSE Challenge Evaluation}
\label{sec:real-avse}

The ISCSLP 2026 Real-World AVSE Challenge adopts a subset of \dataset{} as real-mixture evaluation data~\cite{li2026realavse}. Its Track~1 real-mixture test comprises 2,472 clips and 4,944 target items, versus the full benchmark's 12,076 queries. We evaluate AV-SpatialNet under this protocol, which uses Fun-ASR and target-item mean CER rather than Qwen3-ASR character-micro-average.

\paragraph{System and adaptation.}
The backbone uses the SpatialNet-Large configuration~\cite{quan2024spatialnet}: 12 layers, hidden dimension 192, and feed-forward dimension 384. A single 1-D convolution fuses 1,280-dimensional AV-HuBERT-derived visual features with the real/imaginary mixture STFT channels. Pretraining uses \trainsetname{} training/development data and online three-second reverberant speech mixtures with noise. We then fine-tune using only the challenge development remixes: 900 mixtures and their 1,800 solo recordings. We process the solo recordings with the pretrained AV model to obtain enhanced pseudo-targets for adaptation. The speaker encoder is frozen WeSpeaker ResNet34~\cite{wang2022wespeaker}; Appendix~\ref{sec:real-avse-details} specifies augmentation and optimization.

For enhanced solo pseudo-target $e$, separated waveform $\hat s$, and frozen speaker encoder $E$, fine-tuning minimizes
\begin{equation}
\begin{split}
 \mathcal{L}_{\mathrm{FT}}={}&-\mathrm{SI\mbox{-}SNR}(\hat s,e)\\
 &+w[1-\cos(E(\hat s),E(e))],\quad w=1.
\end{split}
\end{equation}
Unlike L1's margin loss, this aligns outputs with enhanced adaptation-domain references.

\paragraph{Real-mixture test results.}
Table~\ref{tab:real-avse-results} compares the adapted model with audioman and AITD (joint first in official Track~1 ranking) and twilight (third)~\cite{li2026realavse}. Fine-tuning lowers CER from 16.76\% to 12.80\% (3.96 points; 23.63\% relative). The corresponding CER values for the three listed teams are 14.72\%, 18.32\%, and 15.79\%. SPK-SIM improves from 0.5148 to 0.6381 but remains below all three teams; DNS-SIG decreases. Official Track~1 ranks aggregate mix and remix metrics; the table reports a separate real-mixture-subset comparison rather than an official ranking of our model.

\section{Conclusion}

\dataset{} benchmarks both targets of jointly recorded mixtures and contributes a test subset to the Real-World AVSE Challenge. Adapters benefit ResNet; early fusion is stronger with AVSR. Speaker preference and detected overlap can disagree, motivating joint content, identity, and interference diagnostics. VoxBlink2 training improves CER under shared development selection. Separately, challenge adaptation improves CER and speaker similarity while reducing DNS-SIG.

\label{page:main-end}
\clearpage
\section*{Limitations}
\label{sec:limitations}

\paragraph{Coverage and dependence.}
\dataset{} contains read Mandarin from 28 speakers in 14 fixed dyads. Repeated scenes, synchronized views, and paired target queries introduce dependence among observations. The reported results describe this recorded population, primarily in frontal views. Small differences between selected checkpoints should be interpreted in the context of this shared evaluation set.

\paragraph{Automatic diagnostics.}
Speaker margin measures relative embedding preference and can remain positive when both voices are present. JointPass and BothPass combine this preference with an OSD decision at a fixed threshold; they summarize detector outcomes rather than source-reconstruction accuracy. Silence handling follows the rule in Appendix~\ref{sec:metric-details}. CER reflects both the extracted speech and the ASR backend. Scene-matched solo recordings supply identity references, while the overlap stages capture jointly recorded speech.

\paragraph{Comparison scope.}
Visual representations differ in backbone, pretraining, and dimensionality; fusion strategies differ in parameterization and injection location. Corpus comparisons use shared VoxBlink2 development selection. The retained training segments average 10.99 seconds versus 6.06 seconds in the source pool, reflecting the joint validity and quality filters. These comparisons characterize complete configurations under the stated selection protocol. The challenge adaptation study uses its own development remixes and test protocol, with recordings drawn from a subset of the benchmark.

\section*{Ethical Considerations and Data Access}
\label{sec:ethics}

\dataset{} was recorded by a professional data-collection company. All recruited participants signed consent forms covering data collection and public release. \dataset{} is released under \href{https://creativecommons.org/licenses/by-nc/4.0/}{CC BY-NC 4.0}. Speech, video, mouth crops, and identity-linked metadata contain personal information; masking illustrations does not anonymize the underlying recordings. The license permits noncommercial reuse with attribution under its stated conditions.

\trainsetname{} releases file lists, split manifests, and curation instructions under CC BY-NC 4.0. Users obtain authorized access to VoxBlink2 and extract the listed items from the upstream corpus. VoxBlink2 media remain governed by upstream permissions and are not redistributed or relicensed as part of this release. The consent statement above applies to the participants recruited for \dataset{}.

The dataset repositories provide resource descriptions and access information. Benchmark release materials distinguish clip/query manifests, identity-alignment mappings, and original media from derived features. Model configurations and inference instructions specify external visual-frontend dependencies. \datarepositorylinks

\clearpage
\label{page:references-start}

{\sloppy
\bibliography{qiangda-acl-2026}
}

\appendix
\clearpage
\label{page:appendix-start}

\section{Corpus Preparation Details}
\label{sec:corpus-details}

\paragraph{Training-pair validation and quality.}
Candidates from VoxBlink2~\cite{lin2024voxblink2} must have decodable speech, synchronized face tracks, and visible mouth motion. Each retained 16-kHz mono waveform and grayscale $T\times96\times96$ ROI array keeps its identity--video--segment key. The curation implementation scores a 120-band log-power Mel representation with 20-ms windows and 10-ms hops using DNSMOS~\cite{reddy2021dnsmos}; pairs with predicted OVRL below 3.3 are discarded rather than enhanced. Retained recordings can still contain background noise.

Table~\ref{tab:voxblink-curation} summarizes the source and retained subset. Joint validity/quality filtering retains 25.54\% of identities, 2.53\% of utterances, and 4.60\% of duration. The 239,680/11,148 train/dev pairs have disjoint identities. Unlike transcript-oriented LRS3 or large-scale AVSpeech~\cite{afouras2018lrs3,ephrat2018looking}, the subset supplies validated audio/ROI pairs and identity keys for extraction.

\begin{table}[t]
\centering
\footnotesize
\setlength{\tabcolsep}{3pt}
\caption{VoxBlink2 source pool and the curated \trainsetname{} subset. Duration and mean OVRL are computed over retained audio.}
\label{tab:voxblink-curation}
\begin{tabular}{lrr}
\toprule
Metric & VoxBlink2 & \trainsetname{} \\
\midrule
Identities & 111,284 & 28,421 \\
Utterances/pairs & 9,904,382 & 250,828 \\
Duration (h) & 16,672 & 766.17 \\
Mean duration (s) & 6.06 & 10.99 \\
Mean utt./identity & 89.00 & 8.83 \\
Mean DNSMOS OVRL & -- & 3.38 \\
\bottomrule
\end{tabular}
\end{table}

\paragraph{Real-scene acquisition.}
Each scene keeps camera position, lighting, and acoustics fixed across A-only, B-only, and overlap stages. One reader begins during the other's speech. Eye-level central and symmetric side cameras retain both mouths; headset and near-field microphones supplement the camera-synchronized audio. Angle metadata denotes side-camera placement relative to the participant midpoint, not a measured target direction of arrival.

\section{Identity-Alignment Details}
\label{sec:identity-alignment-details}

For every two-person video, adjacent RetinaFace detections are initially associated by bounding-box intersection over union. FAN estimates 68 two-dimensional landmarks; missing detections are filled from neighboring valid landmarks to preserve synchronization. After temporal smoothing, an affine transform based on stable eye and nose points maps each face to a mean-face template before mouth cropping.

For scene-level identity consistency, every third frame is embedded into a 512-dimensional, L2-normalized representation. K-means with $K=2$ initializes two identity prototypes, and the lowest-similarity 15\% of assigned samples are removed before prototype refinement. At sampled time $t$, binary state $z_t$ indicates whether the temporary tracks preserve or swap their order. Given valid tracks $\mathcal{V}_t$, embedding $\mathbf{e}_{t,j}$, prototype $\mathbf{p}_k$, and keep-or-swap permutation $\pi_{z_t}$, we solve
\begin{equation}
\begin{split}
z^*=\arg\min_{\{z_t\}}\;&\sum_t\frac{1}{|\mathcal{V}_t|}
\sum_{j\in\mathcal{V}_t}\left(1-\mathbf{e}_{t,j}^{\top}
\mathbf{p}_{\pi_{z_t}(j)}\right)\\
&+0.35\sum_{t>1}\mathbf{1}(z_t\neq z_{t-1}).
\end{split}
\end{equation}
The fixed transition cost 0.35 penalizes changes in track assignment, balancing framewise identity evidence against temporal continuity.

Frames whose two embeddings have cosine similarity above 0.92 are treated as duplicate detections or track collisions. These frames and unreliable frames near switch boundaries are replaced with neighboring trusted ROIs; face tracks, landmarks, detection flags, and mouth ROIs are reordered together.

Stable tracks do not yet identify the annotation labels. TalkNet scores both tracks for every s1-only and s2-only clip, and the relative activity is computed after discarding the first and last 4\% of frames. The scene-level keep-or-swap decision combines ROC-AUC with the median difference between the two solo-stage score distributions. We use 2,000 bootstrap resamples~\cite{efron1979bootstrap} to assess stability and manually review low-confidence or inconsistent scenes before permanently fixing the person-to-speaker mapping.
\section{Training and Checkpoint Details}
\label{sec:training-provenance}

\begin{table*}[t]
\centering\small
\setlength{\tabcolsep}{7pt}
\caption{Supplementary details for the seven numbered systems. Parameters count the separator backend only, excluding frozen visual frontends and external speaker encoders. Target and Interferer are mean cosine similarities over the same 12,076 outputs as Table~\ref{tab:numbered-configs}.}
\label{tab:numbered-details}
\begin{tabular}{@{}lrrr@{}}
\toprule
ID & Parameters (M) & Target $\uparrow$ & Interferer $\downarrow$ \\
\midrule
B0 & 10.07 & 0.6053 & 0.2901 \\
L1 & 10.07 & 0.6010 & 0.2879 \\
F1 & 11.52 & 0.6231 & 0.2803 \\
B1 & 9.92 & 0.5486 & 0.3322 \\
D1 & 10.07 & 0.6037 & 0.3486 \\
X1 & 11.12 & 0.5622 & 0.3562 \\
D2 & 9.92 & 0.5624 & 0.3603 \\
\bottomrule
\end{tabular}
\end{table*}

Table~\ref{tab:numbered-details} reports separator sizes and component speaker similarities. Models are selected within 50 training epochs; distributed checkpoints retain model tensors but not optimizer/RNG states. The number of optimizer updates within this budget varies with the training corpus.

\subsection{Optimizer and Scheduler}
\label{sec:optimization-details}

FP32 AdamW uses backbone/visual learning rates $10^{-3}/10^{-4}$, weight decays $0/10^{-4}$, betas $(0.9,0.999)$, epsilon $10^{-8}$, and norm-5 clipping. The usual batch is $8\times6$, with smaller final batches allowed. Learning rates halve after three consecutive non-improving validation epochs and on each subsequent non-improving epoch until improvement resets the counter. Early stopping after ten such epochs is available, but all reported runs reach 50 epochs. Development mixtures are sampled online. The entry point seeds Python and PyTorch with 1234. NumPy, DataLoader, and continuation-state coverage remain incomplete in the reproducibility record.

\section{Evaluation Implementation Details}
\label{sec:metric-details}

\paragraph{Text and speaker references.}
All 12,076 outputs, including blank ASR hypotheses, contribute to the 211,866-character CER denominator. Normalization removes enumerated Chinese/ASCII punctuation and whitespace, with no extra simplified/traditional conversion. The speaker evaluator uses 154 scene--speaker reference pools containing 762 solo clips from the 28 participants. The positive-margin boundary is zero, with no additional absolute-similarity cutoff.

\paragraph{OSD edge cases.}
Short outputs are zero-padded to two seconds; a tail-aligned final window is included. The valid-window RMS floor is $\max(-45, R_{\max}-35)$ dBFS, where $R_{\max}$ is the loudest output-window RMS in dBFS. The score averages overlap posteriors across valid windows. Outputs with no valid windows receive zero and pass the overlap criterion by construction; they must still satisfy the positive-margin criterion for JointPass. Thus an overlap pass alone includes low-energy outputs, whereas JointPass also requires relative target-speaker preference.

\paragraph{DNSMOS and reproducibility.}
Non-personalized P.835 scoring~\cite{reddy2021dnsmosp835} uses 16-kHz audio, 9.01-second windows, one-second hops, and polynomial calibration. Short recordings are repeated; scores are averaged within and then across outputs. OSD RMS normalization is not applied. The inference supplement supplies the F1 separator checkpoint, its configuration, and a precomputed-feature inference interface. Visual frontends and evaluators use external weights under their original distribution terms.

\section{Separator Implementation Details}
\label{sec:adapter-details}

Each complete waveform is divided by its standard deviation, floored at $10^{-4}$, with scale restored after separation. The input and transposed-output convolutions use $3\times3$ kernels; the input has 128 channels and group normalization. Intra-frequency and inter-time bidirectional LSTMs have 240 hidden units per direction; attention has four heads. Convolution weights retain ordinary random initialization when the scalar residual coefficient starts at zero.

\paragraph{Low-rank conditioning.}
With rank $r=32$, four bias-free $1\times1$ projections implement
\begin{align}
 a_\ell &= \mathrm{Norm}_{r}(W_{a,\ell}z_\ell),\\
 b_\ell &= \mathrm{Norm}_{r}(W_{v,\ell}\mathrm{LN}(v)),\\
 q_\ell &= W_{u,\ell}\,\mathrm{GELU}\!\left(W_{f,\ell}[b_\ell\odot(1+a_\ell)]\right).
\end{align}
Here $z_\ell\in\mathbb{R}^{B\times128\times T\times65}$ and aligned $v\in\mathbb{R}^{B\times V\times T}$. LN normalizes the $V$ visual channels independently at each frame, without affine parameters. Likewise, $\mathrm{Norm}_r(h)=(h-\mu_r)/\sqrt{\sigma_r^2+10^{-6}}$ uses the population mean and variance over the $r$ bottleneck channels only: separately at each time--frequency location for audio and each frame for video, without affine parameters. Normalization statistics are computed in FP32. The projected visual term has shape $B\times32\times T\times1$ and broadcasts over frequency; its interaction with $a_\ell$ makes the residual frequency-dependent. Define $\mathrm{RMS}_{\epsilon}(h)=\sqrt{\mathrm{mean}_{C,F}(h^2)+\epsilon}$, with $\epsilon=10^{-6}$. Direction and magnitude are controlled separately:
\begin{align}
 d_\ell &= q_\ell/\mathrm{RMS}_{\epsilon}(q_\ell),\\
 U_\ell &= \alpha_\ell\,\mathrm{stopgrad}(\mathrm{RMS}_{\epsilon}(z_\ell))d_\ell,\\
 \alpha_\ell &= 0.2\,\mathrm{softsign}(s_\ell),\qquad s_\ell^{(0)}=0.
\end{align}
The update starts at zero and is bounded relative to frame-level audio RMS, not elementwise clipped. The fusion comparison evaluates this internal residual interface against input-level conditioning, varying both injection location and parameterization.

\section{Real-AVSE Reproduction Details}
\label{sec:real-avse-details}

\paragraph{Frontend and augmentation.}
The experiment record cites the multi-view encoder of \citet{su2025enhanced} for the 1,280-dimensional visual representation. Three-second online mixtures use target/interferer speech from \trainsetname{}, DEMAND OHALLWAY\_16k noise~\cite{thiemann2013demand}, and FRAM-RIR~\cite{luo2024framrir}. Room sampling follows the cited reverberant augmentation setup~\cite{ko2017reverberant}: small/medium/large probabilities are 0.8/0.1/0.1; absorption coefficients are drawn from three ranges spanning $[0.2,0.8]$ with probabilities 0.3/0.4/0.3.

The recorded nominal draws are $\mathrm{SNR}=\max(\mathcal{N}(10,4),5)$ dB and $\mathrm{SIR}=\max(\mathcal{N}(5,10),-5)$ dB. Mixtures follow $x=s_t*h_t+\alpha(s_i*h_i)+\beta n$, with $\alpha=10^{-\mathrm{SIR}/20}$ and $\beta=10^{-\mathrm{SNR}/20}$. The document leaves Gaussian second-argument conventions, component-power normalization, room-size boundaries, and absorption-bin edges unspecified; these details are needed to recover the realized mixture SIR/SNR.

\paragraph{Optimization and evaluation.}
Both stages use Adam at initial rate $10^{-3}$, halved after three non-improving validation checks, with stopping after ten. Pretraining minimizes negative SI-SNR; adaptation uses the main-text pseudo-target objective. Solo recordings provide remix sources but retain room and recording noise. Enhancing them before supervision is intended to reduce these components rather than reproduce them. The resulting supervision follows the pretrained model's estimate of the solo speech. The challenge uses Fun-ASR and target-item averaging, not the numbered experiments' Qwen3-ASR character-micro-average~\cite{li2026realavse}. Table~\ref{tab:real-avse-results} reports the real-mixture-subset scores; the official ranks shown for the competing teams refer to the challenge's combined mix/remix protocol.

The challenge overview reports audioman and AITD as joint first and twilight as third on Track~1. It describes AITD development-set fine-tuning but provides no twilight system description. The statement that all three comparators also use remix fine-tuning follows the supplied comparison information; detailed recipes are not independently matched. The 2,472-clip challenge test shares recordings with the benchmark and provides a complementary evaluation protocol. The challenge reports seven dyads split three/four between development/test; a scene- and speaker-level crosswalk to all 14 benchmark dyads is not yet supplied. Acoustic mismatch between enhanced pseudo-targets and speaker references remains an untested explanation for SPK-SIM.

\begin{figure*}[!p]
\section{Analysis of the Seven Selected Systems}
\label{sec:additional-analysis}

\begin{minipage}[t]{0.48\textwidth}
\paragraph{Content and joint pass.}
Figure~\ref{fig:final-system-analysis}(a) compares CER and JointPass for the seven final systems. F1 achieves the lowest CER (22.99\%), while D1 has the highest JointPass (48.37\%) at a higher CER of 34.98\%. Adding the auxiliary speaker loss (B0--L1) likewise increases both CER and JointPass.
Comparing F1 with D1 shows that stronger identity/overlap scores can accompany less accurate target transcription. Reporting both axes keeps this trade-off visible rather than reducing the benchmark to a single ranking.
\end{minipage}\hfill
\begin{minipage}[t]{0.48\textwidth}
\paragraph{Identity preference and overlap.}
Figure~\ref{fig:final-system-analysis}(b) relates positive speaker margin to detected residual overlap. F1 has the highest positive-margin rate (93.38\%), yet 55.45\% of its outputs exceed the overlap threshold. B0 likewise combines a 90.68\% positive-margin rate with 75.11\% detected overlap. Relative identity preference and the overlap diagnostic capture distinct aspects of the outputs.
\end{minipage}
\par\medskip
\centering
\includegraphics[width=\textwidth]{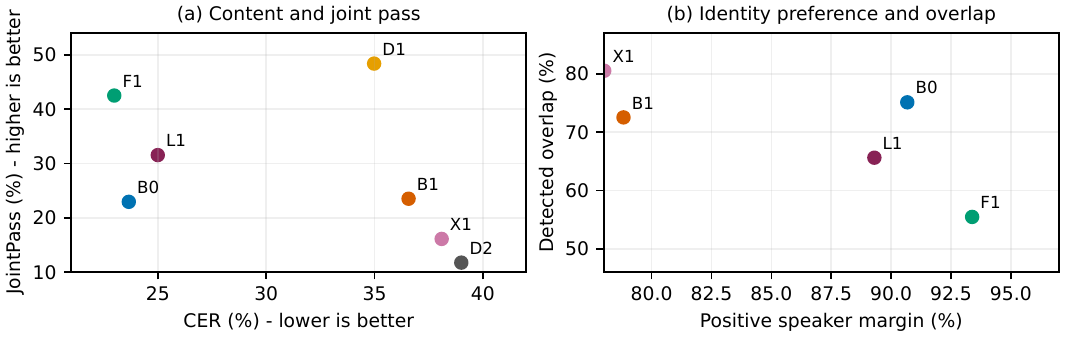}
\caption{Metric relationships for the seven selected final systems, each evaluated on all 12,076 target queries. (a) CER versus JointPass; (b) positive speaker-margin rate versus detected residual overlap. Labels identify the same checkpoints as the main tables; percentages are reproduced from their evaluation records.}
\label{fig:final-system-analysis}
\par\medskip
\begin{minipage}[t]{0.48\textwidth}
\paragraph{Paired-target outcomes.}
Figure~\ref{fig:paired-target-outcomes} groups each mixture by the number of target outputs satisfying JointPass. F1 passes for both targets on 20.77\% of mixtures and for exactly one on 43.44\%; D1 reaches 26.55\% and 43.64\%, respectively. Their exactly-one fractions differ by only 0.20 percentage points, whereas D1's both-target fraction is 5.78 points higher. D1's higher pass rate thus includes a larger both-target fraction.
\end{minipage}\hfill
\begin{minipage}[t]{0.48\textwidth}
\paragraph{Fusion and paired consistency.}
For ResNet conditioning, B1 increases the both-target fraction from X1's 3.23\% to 5.68\% and reduces the neither-target fraction from 71.05\% to 58.69\%. Most of that distributional shift appears in the exactly-one category (25.72\% to 35.62\%). Across all seven systems, exactly-one passing is more frequent than both-target passing: a favorable result for one query often coexists with a failed paired query.
\end{minipage}
\par\smallskip
\includegraphics[width=\textwidth]{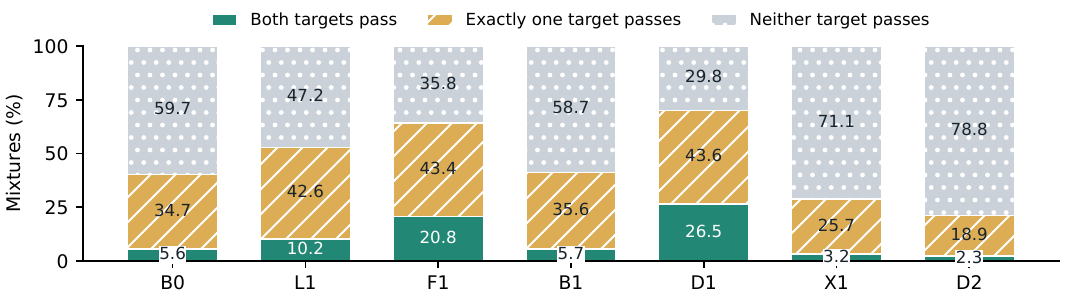}
\caption{Paired-target diagnostic outcomes on the same 6,038 mixtures for each final system. A target passes when its speaker margin is positive and its OSD score satisfies the overlap criterion. For output-level JointPass $J$ and mixture-level BothPass $B$, the three fractions are $B$, $2(J-B)$, and $1-2J+B$. Counts are recovered from the unrounded evaluation rates; labels show percentages.}
\label{fig:paired-target-outcomes}
\end{figure*}

\begin{figure*}[!p]
\section{A Paired-Target Extraction Case}
\label{sec:paired-case}
\begin{minipage}[t]{0.48\textwidth}
\paragraph{Setup and selection.}
Figure~\ref{fig:paired-target-case} shows a four-second real mixture, \texttt{r2m\_s042\_c00025}, recorded at 0.5\,m. The final B1 and X1 checkpoints each process the same waveform twice, changing only the target mouth sequence. Both systems use frozen ResNet-512 features, with adapter conditioning in B1 and early fusion in X1. Each target has 100 mouth frames aligned with 64,000 audio samples at 16\,kHz. We selected this illustration by listening to six near-field candidates alongside same-scene solo references.

\paragraph{Rendering.}
Full-utterance FP32 inference uses the final weights and identical feature preprocessing. Spectra use equal waveform RMS, a 32-ms Hann window, an 8-ms hop, and a shared color scale.

\end{minipage}\hfill
\begin{minipage}[t]{0.48\textwidth}
\paragraph{Two targets from one mixture.}
Listening inspection indicates that both systems largely recover the respective target speakers when the visual query is switched. Successful conditioning preserves a different target speaker for each query while the acoustic input remains unchanged. This example illustrates successful behavior with either fusion strategy; the differences in their overall reliability are captured by the full-set paired outcomes in Figure~\ref{fig:paired-target-outcomes}. The reference recordings support an informal voice comparison and contain different utterances, rather than synchronized clean sources for this mixture.

\end{minipage}
\par\medskip
\centering
\includegraphics[width=\textwidth]{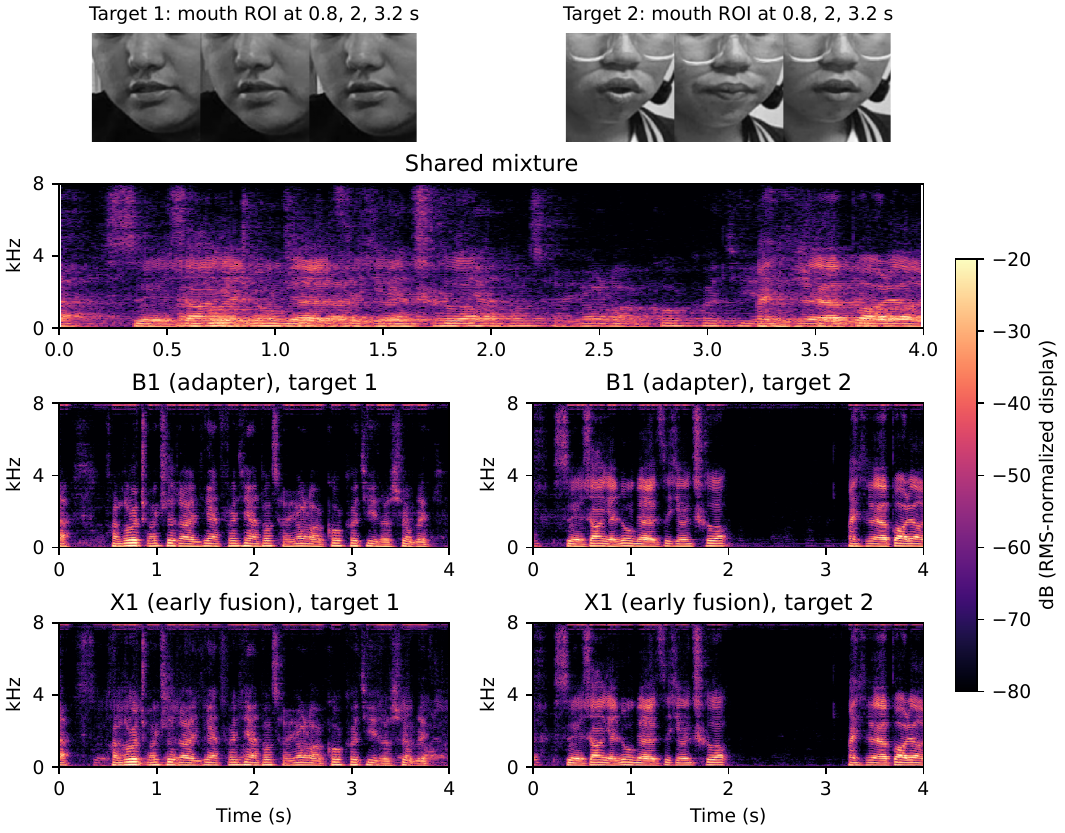}
\caption{A listening-selected paired-target example from \dataset{}. Top: three mouth-ROI frames per target. Below: the shared mixture and outputs conditioned on target 1 or target 2 from B1 (adapter) and X1 (early fusion). Listening inspection finds both systems largely recover the intended speakers in this example. All spectra use equal RMS normalization and a shared color scale.}
\label{fig:paired-target-case}
\end{figure*}

\end{document}